\documentclass[sigconf]{acmart}

\AtBeginDocument{%
	\providecommand\BibTeX{{%
			\normalfont B\kern-0.5em{\scshape i\kern-0.25em b}\kern-0.8em\TeX}}}

\newcommand{\figref}[1]{Fig.~\ref{#1}}
\newcommand{\secref}[1]{Sec.~\ref{#1}}
\copyrightyear{2019} 
\acmYear{2019} 
\setcopyright{rightsretained} 
\acmConference[DEBS '19]{DEBS '19: The 13th ACM International Conference on Distributed and Event-based Systems}{June 24--28, 2019}{Darmstadt, Germany}
\acmBooktitle{DEBS '19: The 13th ACM International Conference on Distributed and Event-based Systems (DEBS '19), June 24--28, 2019, Darmstadt, Germany}
\acmDOI{10.1145/3328905.3332513}
\acmISBN{978-1-4503-6794-3/19/06}

\begin{document}

%
\title[Infrastructure fo Processing Financial Data]{Poster: A Real-World Distributed Infrastructure for Processing Financial Data at Scale}
\titlenote{This is the author's version of the accepted work posted here for your personal use. Not for redistribution. Definitive Version of Record published in {DEBS '19} (see below).}

\author{Sebastian Frischbier}
\author{Mario Paic}
\author{Alexander Echler}
\author{Christian Roth}
\affiliation{%
	\institution{vwd: Vereinigte Wirtschaftsdienste GmbH}
	\country{Germany}
}

\renewcommand{\shortauthors}{Frischbier et al.}

\begin{abstract}
Financial markets are event- and data-driven to an extremely high degree. For making decisions and triggering actions stakeholders require notifications about significant events and reliable background information that meet their individual requirements in terms of timeliness, accuracy, and completeness. 
As one of Europe’s leading providers of financial data and regulatory solutions vwd: processes an average of 18 billion event notifications from 500+ data sources for 30 million symbols per day. Our large-scale distributed event-based systems handle daily peak rates of 1+ million event notifications per second and additional load generated by singular pivotal events with global impact.

In this poster we give practical insights into our IT systems. We outline the infrastructure we operate and the event-driven architecture we apply at vwd. In particular we showcase the (geo)distributed publish/subscribe broker network we operate across locations and countries to provide market data to our customers with varying quality of information (QoI) properties.
\end{abstract}

%



%

%
\maketitle

%
%

\section{Introduction}

Having access to reliable, accurate, fresh, and complete information about financial markets is essential for stakeholders. Raw market data about financial instruments (e.g., stocks/securities, funds) and their instances (symbols; e.g., Apple Inc.) is provided as structured and unstructured data by various sources such as exchanges, financial institutions, and rating agencies. Financial data vendors and solution providers, such as vwd, collect this data, purge, and enrich it before providing the resulting condensed information together with additional key performance indicators (KPIs) to subscribers based on their quality of information (QoI) requirements. 

vwd operates an extensive geo-distributed infrastructure and applies an event-driven architecture while meeting restrictive requirements in terms of availability, performance, and IT compliance. We use (event) stream-processing and complex event-processing (CEP) to derive various KPIs and regulatory solutions on the fly from our real-time data streams and historic reference data. 

In this contribution we sketch out our infrastructure (\secref{sec:infrastructure}) and the event-driven architecture (\secref{sec:architecture}) of our ticker plant (\secref{sec:feedhandling}) before showcasing our (geo)distributed Pub/Sub broker network we run across locations to deliver fine-granular market data (\secref{sec:pubsubbrokers}). We conclude our contribution by outlining ongoing work (\secref{sec:outlook}).

\section{vwd's Physical Infrastructure}
\label{sec:infrastructure}

We rely on a hybrid infrastructure that is owned and operated by us for most parts. 
On the physical level, we operate out of multiple dedicated data centers, collocation sites, managed hosting environments, and public cloud platforms. Due to compliance and local jurisdiction we store customer data only at sites across continental Europe but are also present with our own hardware on-site at local exchanges such as Hong Kong or London. Main drivers for our redundant geo-distributed setup are legal and regulatory constraints (e.g., data locality), disaster resilience, and localized products.

\begin{figure}[!ht]
	\centering
	\includegraphics[width=0.45\textwidth]{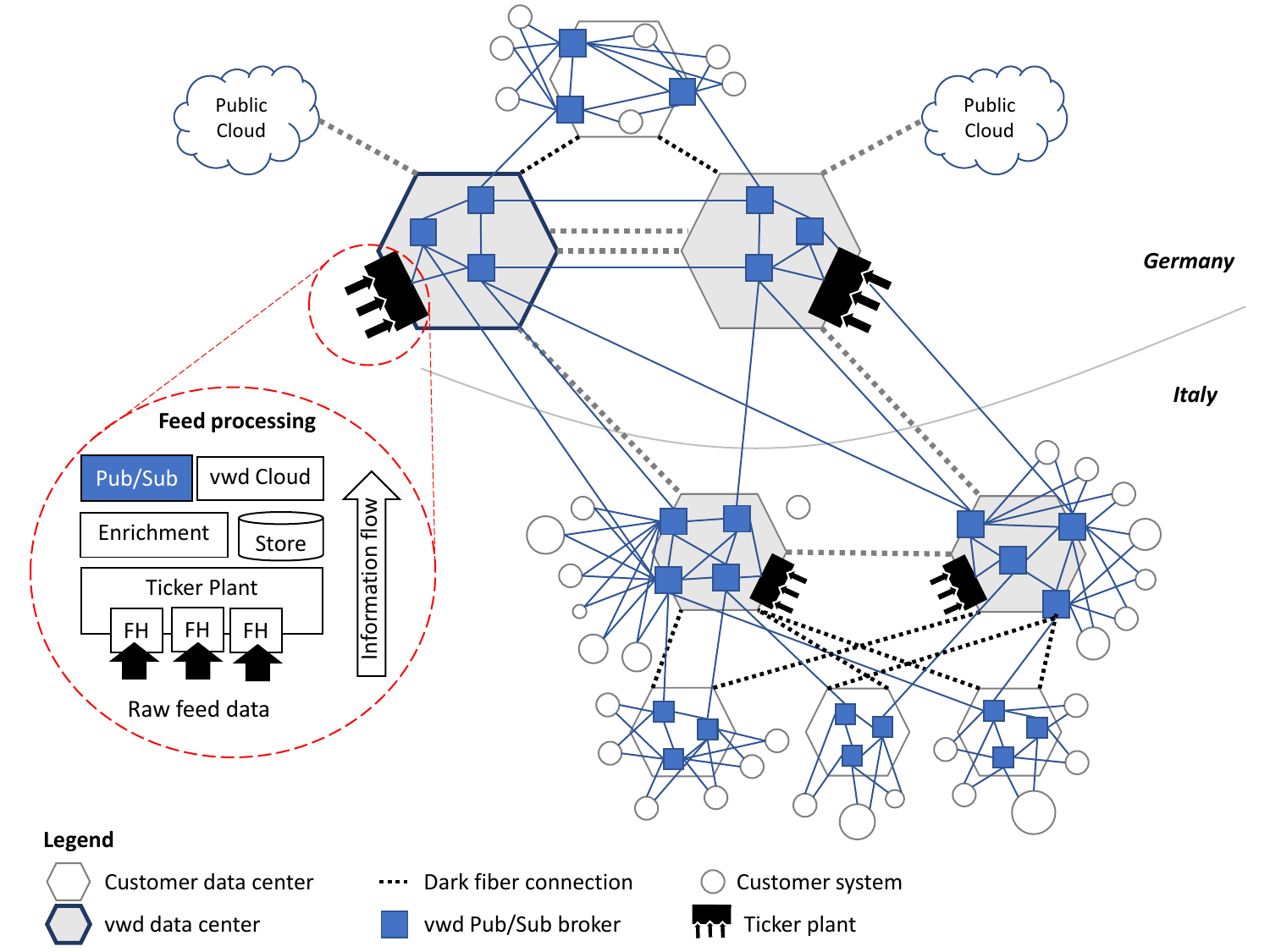}
	\caption{Simplified subset of vwd's geo-distributed infrastructure (here: Germany and Italy): logical links of meshed distributed Pub/Sub broker network (blue) and details on feed handling (red). Data contributors omitted.}
	\label{fig:infrastructure}
\end{figure}

\figref{fig:infrastructure} illustrates a subset of our physical infrastructure (here: Germany and Italy) that we will use as a running example. All our sites are interconnected with redundant dedicated gigabit lines (dark fiber, dotted black lines) that also connect us with most of our data contributors (omitted in the figure) and customers. While some data feeds are consumed and provided via public internet, using redundant dedicated lines is still the predominant mode for exchanging financial data streams due to their volume and latency requirements. Similarly, we use dedicated connections from our own data centers to entry points of global public cloud providers to utilize their resources transparently within our network. At the time of writing, one such connection provides 10~Gbit bandwidth; multiple connections can be trunked for an increased limit. The setup of our central data center sites is designed for failover and disaster recovery scenarios so that one unavailable site can be compensated by another. For our internal backbones we also rely on gigabit network and software defined networks (SDN): 10~Gbit for traffic-intensive platforms and 1~Gbit for less utilized environments. 

\section{Event-driven Architecture} 
\label{sec:architecture}
In our group-wide architecture we combine the two complimentary paradigms of event-processing and service-oriented architectures (SOA). 
Close to the data sources, event-driven feed processing systems are tightly integrated vertically with our infrastructure as we have to customize network protocols and align software with hardware to achieve maximum processing performance. For the customer-facing parts of our architecture (vwd Cloud), in turn, we focus on decoupling systems and infrastructure by using open and commercial platforms, virtualization, and (micro)service architectures based on Docker (Swarm) and Kubernetes.

\subsection{Feed Processing: Ticker Plant}
\label{sec:feedhandling}
As shown in the exploded view (red) on the left hand side of \figref{fig:infrastructure}, \emph{feed handlers} (FH) in the \emph{ticker plant} of our central processing system are subscribed to our various data sources to receive, check, purge and normalize the incoming feeds. Each feed handler is tailored to match the specific protocol, syntax, and semantics of its feed(s). For some feeds, basic aggregations for composite events are already computed by the feed handlers. Frequent basic enrichment we do here is adding standard properties if they are not available in the original notification. Examples are total volume, open/close, or day high/low. They are calculated in real-time and appended to the incomplete notification instead of creating a separate one.  The normalized data is fed into our \emph{Event Store} together with other reference data provided by customers or contributors. Derived / complex events are detected in the \emph{Enrichment} component. They are provided to our data-driven solutions, pushed as enriched feeds to our subscribers, and stored in the event store. Here, the normalised data is further enriched with historic / contextual data.

\subsection{Distributed Pub/Sub Broker Network} 
\label{sec:pubsubbrokers}
Market data is fuelling a wide spectrum of applications but in particular market data terminals. Customers using such terminal solutions with high resolution market data usually value timeliness over completeness, i.e., they prefer single notifications to be dropped if the information displayed on the terminal therefore always reflects the most recent events; customers using the same data for analytics usually prefer completeness over timeliness. Consequently, a pub/sub infrastructure must support varying levels of QoI~\cite{frischbier2014managing}.

For pushing our market data from our ticker plants to consuming applications we operate a \emph{content-based distributed publish/subscribe broker network}. As illustrated by the blue overlay network in \figref{fig:infrastructure}, this decentralized broker network runs geographically distributed across multiple data centers and inter-company-borders, i.e., some brokers run on-premise in locations of our customers. All brokers are intermeshed for fault-tolerance and disaster resilience. Apart from our market data universe about 30 million symbols we do also distribute news, alerts, internal system notifications (e.g., configuration changes), and monitoring metrics through the same broker network. The implementation is closed-source.

The main reason for us to rely on a distributed broker network instead of a centralized messaging solution is to combine disaster resilience and compliance with optimized traffic shaping and horizontal scalability. Our broker network mirrors the topology of our physical infrastructure and avoids a single point of failure. Through subscription filtering based on source, type/instrument, or even symbol (i.e., sub-instrument level) together with QoI properties we minimize traffic across locations while maximizing fan-out within a data center through application-level multicast~\cite{frischbier2013aggregation}. We apply a content-based subscription model as we support a huge number of instruments and allow for symbol-level subscriptions -- a topic-based approach would be unmanageable in our case.

\section{Conclusion and Outlook}
\label{sec:outlook}

Within the limits of a poster we outlined the physical infrastructure and event-driven architecture used by vwd to provide financial data at scale. In particular we showcased the (geo)distributed Pub/Sub broker network we use to push time-critical data to our customers.

As part of ongoing work we evaluate open platforms such as Kafka, NATS, or STORM to replace in-house components in our event-driven architecture; in parallel we increase the level of virtualization in our ticker plant on host- and application-level (i.e., containerisation) to benefit from increased portability and scalability as we already do in our vwd Cloud.

\bibliographystyle{ACM-Reference-Format}
\bibliography{InfrastructureForProcessingFinancialDataAtScale-DEBS2019-CR-authorscopy}

\end{document}